# Bulk crystal growth and electronic characterization of the 3D Dirac Semimetal Na$_3$Bi

Satya K. Kushwaha[1], Jason W. Krizan[1], Benjamin E. Feldman[2], András Gyenis[2], Mallika T. Randeria[2], Jun Xiong[2], Su-Yang Xu[2], Nasser Alidoust[2], Ilya Belopolski[2], Tian Liang[2], M. Zahid Hasan[2], N. P. Ong[2], A. Yazdani[2] and R. J. Cava[1]

[1]Department of Chemistry, Princeton University, Princeton NJ 08542

[2]Department of Physics, Princeton University, Princeton NJ 08542

High quality hexagon plate-like Na$_3$Bi crystals with large (001) plane surfaces were grown from a molten Na flux. The freshly cleaved crystals were analyzed by low temperature scanning tunneling microscopy (STM) and angle-resolved photoemission spectroscopy (ARPES), allowing for the characterization of the three-dimensional (3D) Dirac semimetal (TDS) behavior and the observation of the topological surface states. Landau levels (LL) were observed, and the energy-momentum relations exhibited a linear dispersion relationship, characteristic of the 3D TDS nature of Na$_3$Bi. In transport measurements on Na$_3$Bi crystals the linear magnetoresistance and Shubnikov-de Haas (SdH) quantum oscillations are observed for the first time.

**Introduction**

The growth and study of topological materials, which possess remarkable electronic properties, has recently developed into an active area of research [1-8]. The recent discoveries that native Na$_3$Bi and Cd$_3$As$_2$ are three dimensional (3D) Dirac semimetals represent a significant advance in this field, because they enable the study of a new type of topological material [9-12]. Both of these compounds possess bulk Dirac fermions in 3D at the Fermi level that disperse linearly along all three momentum directions, in contrast to the two-dimensional Dirac fermions present on the surfaces of 3D topological insulators and in graphene. The 3D bulk Dirac point is protected by crystal symmetry and contains two overlapping (degenerate) bands in momentum space. These materials are precursors to the so-called Weyl semimetals (WSMs), which may be realized by applying a magnetic field in certain crystallographic directions to split the Dirac point into two Weyl nodes [11,13-16]. Recent theoretical studies also predict the presence of surface states in Dirac semimetals [9,11,17]. Angular-resolved photoemission spectroscopy (ARPES) of the band dispersion, and scanning tunneling microscopy (STM) studies of the Landau quantization and quasiparticle interference have provided evidence that bulk $Cd_3As_2$ is a 3D Dirac semimetal, and its high electronic mobility has been evidenced by transport studies [18-22].

The Bismuth square net compounds $SrMnBi_2$ and $LaAgBi_2$ are also reported to exhibit bulk Dirac fermions [23-25]. In these materials, the bulk Dirac fermions are not protected by crystal symmetry and can be gapped, and they also exhibit quasi 2D Fermi surfaces. $Na_3Bi$ and $Cd_3As_2$ are different, however, because in these two materials the Dirac fermions are stabilized by the crystal symmetry. $Na_3Bi$ and $Cd_3As_2$ can also be classified as "topological 3D Dirac semimetals"; when time-reversal or inversion symmetry is broken in $Na_3Bi$ and $Cd_3As_2$, they become Weyl semimetals. ARPES studies specifying the Dirac semimetal nature of $Na_3Bi$ have also been reported [10], but as yet no electronic transport measurements have been reported, nor have STM measurements exploring the local electronic behavior.

The $Na_3Bi$ crystals used for published experimental (ARPES) investigations were grown by slowly cooling a stoichiometric melt, and were said to have Na point defects from vacancies and Na migration [10]. The crystallization of $Na_3Bi$ from a stoichiometric composition melt is nontrivial. The phase diagram shows that a range of nonstoichiometric Na-Bi compositions is stable for "$Na_3Bi$" near the melting point. The exact melting relations have not been determined, but "$Na_3Bi$" melted at high temperatures in our experience always shows the presence of NaBi as an impurity. Thus for crystals of "$Na_3Bi$" made from slow-cooling nominally stoichiometric melts, NaBi is frequently found intergrown with the crystals. NaBi is a bulk superconductor [26], and interferes with transport and other nonlocal measurements on such samples. Recently, $Na_3Bi$ epitaxial thin films grown by molecular beam epitaxy (MBE) have been reported to display a significant number of Na vacancies as well as *p*-type behavior [27,28]. It is well known that the composition and quality of crystals play a vital role in realizing the optimal electronic properties of materials [29], so obtaining pure Na$_3$Bi crystals and improving their quality is crucial for the optimal characterization of their unique electronic properties. The unusual thermodynamics of crystallization and non-stoichiometry, coupled with the extreme air sensitivity of Na$_3$Bi, present significant experimental challenges in the growth and characterization of high quality single crystals. Na3Bi crystallizes in the space group P6$_3$/mmc with $a$ = 5.448 Å and $c$ = 9.655 Å. It has a simple crystal structure with three distinct crystallographic sites: Na(1), Na(2) and Bi. It consists of Na(1)-Bi honeycomb layers stacked along the $c$ axis, with Na(2) between the layers (Fig. 1). This bulk crystal symmetry leads to the protection of the 3D Dirac fermions.

Realizing the importance of Na$_3$Bi crystals for topological physics, here we report the crystal growth of pure Na$_3$Bi bulk single crystals from an Na flux and a study of their electronic properties. Comparison with crystals grown from a stoichiometric melt is presented. The crystal structure and phase purity of crystals are analyzed by X-ray diffraction, and a low temperature STM was used to study the crystalline quality and 3D Dirac semimetal behavior. Electronic transport measurements were also performed to study the bulk quantum oscillations and magnetoresistance of Na$_3$Bi. The electronic band structures and surface states of each type of crystal were generally characterized by ARPES measurements. The details of the observation of Fermi arcs on the surfaces of these Na-flux grown crystals of Na$_3$Bi are described elsewhere [30].

**Experimental**

The Na-Bi phase diagram ([31]) has a broad region of stable compositions at high temperatures centered around the stoichiometric Na$_3$Bi ratio [Fig. 1]. This broad, oval nonstoichiometric region persists down to temperatures of 370 °C, which leads to many experimental challenges in the crystal growth. Thus different crystallization routes lead to dramatic variability of both crystal quality and electronic properties, especially those related to defect concentration, as "Na$_3$Bi" may be both Na-rich and Na-deficient. Here we report the growth of crystals from both stoichiometric and off-stoichiometric (Na-flux) compositions.

For the stoichiometric-melt cooled crystallization of Na$_3$Bi, a 3:1 atomic ratio [31] of Na (99.5%) and Bi (99.999%) was used. The samples were prepared in an Ar filled glove box, with residual oxygen and water below the detection limit of 0.1 ppm. The samples were placed in one-end-sealed tubes of either C-coated stainless steel (304 L) or electrical grade Cu. To carbon coat the tubes, they were heated in an rf (radio frequency) furnace under the continuous flow of argon (Ar) and a small amount of acetone, which thermally breaks down, was added when they were hot. The open ends of the tubes after starting material insertion were crimped and welded under a continuous Ar-flow. The sealed ampoules were heated to 900 °C at 60 °C/h for 6 hrs, cooled to 750 °C at 10 °C/h and further cooled to 300 °C at 1.5 °C/h, after which they were left to anneal at 300 °C for 24 hrs and finally furnace cooled to room temperature. We found that better crystals were formed in the stainless steel tubes, which are designated as "$Na_3Bi(1)$".

For the Na-flux grown crystals, a Na rich composition [Fig. 1] of Na plus Bi in a 90:10 atomic percent ratio was chosen. The starting materials were placed in C-coated steel tubes and steel wool was used as a plug. The sealed ampoules were heated to 850 °C at a rate of 60 °C/h for 12 hours to ensure homogeneous mixing of the elements, and then cooled to 750 °C at 10 °C/h. For the entire process of nucleation and growth the temperature was lowered slowly to 325 °C, at a rate of 1.5 °C/h. The ampoules were then centrifuged while hot and designated as "$Na_3Bi(2)$". Since $Na_3Bi$ crystals are extremely air sensitive, they were harvested and stored in a glove box.

To confirm the crystal structure and assess the phase purity and crystalline quality, a powdered specimen of $Na_3Bi(1)$ and a well faceted single crystal plate of $Na_3Bi(2)$ were studied in a Bruker D8 FOCUS X-ray diffractometer, with a Cu Kα radiation and a graphite diffracted beam monochromator. To avoid exposure to air, the specimens were covered with a thin layer of Paratone-N oil followed by a layer of Kapton foil. The diffraction patterns were recorded in the 2θ angular range of 10–90 degrees with a step rate of 0.04°/s. To investigate the electronic properties of $Na_3Bi$, we used a home built low-temperature STM to obtain topographic and spectroscopic information with high spatial resolution [18, 32]. The $Na_3Bi$ crystals were mounted on a stage in the glove box and transferred to the STM systems. The STM measurements were performed at a temperature of approximately 3K. A similar procedure was employed for preparing the samples for transport study.

Pt wire contacts were mounted to the sample using silver epoxy inside the clean Ar glove box environment. The samples were heated in the glove box to 120 °C to cure the epoxy. Afterwards, the sample was immersed in Paratone-N oil and immediately transferred to the cryostat for transport measurements, without exposure to air. The insert was cooled down to 4 K immediately. The resistance was measured by a Linear Research Resistance Bridge.

Angle-resolved photoemission spectroscopy (ARPES) experiments were performed at beamline 5-4 at the Stanford Synchrotron Radiation Lightsource (SSRL) at the Stanford Linear Accelerator Center in California, USA and beamline I4 at the MAX-lab in Lund, Sweden. The energy and momentum resolution was better than 30 meV and 1% of the surface Brillouin zone (BZ) for spin-integrated ARPES measurements. Samples were measured under a vacuum condition of better than $1 \times 10^{-10}$ torr at all beamlines, at liquid nitrogen temperature at the I4 beamtime at the MAX-lab, and at 10-20 K at the beamline 5-4 at the SSRL. Since $Na_3Bi$ is air sensitive, argon-filled glove boxes were used in the entire preparation process.

**Results and discussion**

Figures 1(a) and (b) show typical $Na_3Bi$ specimens grown by each method. The crystals have a metallic purple luster and are highly air sensitive. The stoichiometric-melt cooled crystals are agglomerations of multiple crystal domains and cleave easily, producing shiny surfaces. The flux-grown crystals possess a well-defined hexagonal faceted morphology with a lateral size of up to about 5 mm and thickness around 0.5 mm. The crystal morphology of the flux-grown crystals suggests that the [001] direction is the slowest to grow, resulting in larger *c*-planes. Figure 1(d) depicts a schematic representation of the physical morphology of the hexagonal crystals, with an arrow indicating the [001] direction normal to the (001) planes. The flux grown crystals are easy to cleave parallel to both the (001) and (100) planes. Figure 1(c) shows a schematic of the hexagonal unit cell of the honeycomb crystal lattice [16]. Figure 1(e) shows the diffractograms for typical specimens; $Na_3Bi$(1) powder and $Na_3Bi$(2) single crystal. The top pattern (blue) is for a powder specimen of crystals obtained from the slow cooled stoichiometric melt. The pattern consists primarily of the peaks corresponding to those of $Na_3Bi$; the peak positions and their intensity perfectly match the red vertical lines, which are the data reported in the ICSD [33]. The peaks of the powder pattern are also quite sharp, evidence for high crystal quality. In addition to the main $Na_3Bi$ phase, further peaks are observed in the powder patterns of stoichiometric-melt grown crystals. The additional peaks, indicated by red stars, reveal the presence of NaBi as an impurity intergrown with the $Na_3Bi$. We interpret this to indicate that either the highest melting composition in the "$Na_3Bi$" solid phase field is Na-rich, or that there is inevitably Na loss from stoichiometric melts due to Na's extreme reactivity at high temperatures towards any impurity elements present. The bottom pattern (orange) is recorded in a diffraction geometry where the perpendicular to the plane of the large face of a flux-grown crystals bisects 2θ (Fig. 1(b)). It consists of the peaks corresponding to only the (00l) reflections, confirming that the larger faces correspond to the (001) planes. The diffraction peaks are very sharp, indicating the high quality of the grown single crystals. We also recorded the powder pattern of ground flux grown single crystals (not shown), which agreed well with the Na3Bi pattern in the ICSD.

We have studied multiple samples of $Na_3Bi$ in the STM that were cleaved in ultra-high vacuum immediately prior to measurement. Two distinct surface morphologies were observed. Data measured on both surfaces are characteristic of a linear energy-momentum dispersion relation, as expected for a 3D Dirac semimetal. We first describe measurements on a $Na_3Bi$(1) crystal grown by cooling a stoichiometric melt. A topographic image of a single terrace on the cleaved surface of this sample is shown in Fig. 2(a), and a higher resolution image is presented in Fig. 2(b). The surface topography is dominated by one-dimensional (1D) lines separated by approximately 9.1 Å. These features are prominent in the discrete Fourier transform shown in the inset of Fig. 2(a), which also reveals a faint hexagon of points separated by about 0.8 Å$^{-1}$ from the origin that result from the quasi-periodic bright spots along the 1D lines in real space. Figure 2(c) shows a typical line cut spanning multiple terraces, showing an average step height for this cleavage plane of 4.6 Å. This step height is characteristic of the distance between (100) planes in $Na_3Bi$. The 9.1 Å in-plane line spacing is close to the *c*-axis lattice constant of $Na_3Bi$, but we are unable to resolve a fully periodic lattice structure in the image and therefore cannot uniquely identify the cleavage plane or determine whether the surface is reconstructed. The exact crystallographic character of this surface is not critical for the analysis performed below.

To probe the band structure of $Na_3Bi$, we measured the differential conductance *dI/dV* as a function of the perpendicular magnetic field *B* and the bias *V* between tip and sample on this surface. For a typical metallic tip whose density of states is independent of energy, *dI/dV* is proportional to the local density of states (LDOS). Applying a magnetic field quantizes the electron motion perpendicular to the field direction but does not affect the momentum in the parallel direction. This yields the 3D analogue of Landau levels (LLs), and peaks in the density of states are expected to occur any time there is a van Hove singularity caused by a locally flat region of the dispersion parallel to the applied field. For 3D Dirac materials, van Hove singularities are predicted near the Γ point for all LLs with nonzero orbital index $N$, and the corresponding LL energies are expected to scale as $E_N = E_{DP} \pm v_F\sqrt{2e\hbar B|N|}$, where $e$ is the electron charge, $\hbar$ is Planck's constant divided by $2\pi$, $E_{DP}$ is the Dirac point energy and $v_F$ is an effective Fermi velocity averaged over the directions perpendicular to the magnetic field [34]. Figure 2(d) shows $dI/dV$ at several magnetic fields for a single location on the sample. Weak oscillations are visible in the density of states, and the peaks disperse as the magnetic field is changed. We therefore attribute these features to 3D LLs in Na$_3$Bi. The oscillations are only visible at some locations on the surface, likely due to defects in the underlying crystal, and, as was seen in Cd$_3$As$_2$, we only observe LLs on the electron side of the Dirac point [18]. Nonetheless, we can assign an orbital index to each peak and plot the energy dependence of these features as a function of wavevector $k_N = \sqrt{2eB|N|/\hbar}$ (Fig. 2(e)). A linear dispersion is clearly seen, as expected for a 3D Dirac semimetal. From this plot, we extract $v_F = 5.22$ eVÅ from the fitted slope as well as $E_{DP} = -71$ meV, showing that this sample is *n*-doped.

Measuring spatial variations in the LDOS with a STM can also provide information about the electronic band structure of a material. Elastic scattering between points on a constant-energy contour leads to a standing wave pattern of modulated LDOS (known as quasiparticle interference, or QPI), whose Fourier transform then shows the scattering wavevectors that occur at that energy. For the above Na$_3$Bi(1) sample, we observe signatures of QPI (Fig. 2(f-h)) at multiple energies, and in all cases, their Fourier transform shows a circular pattern (Fig. 2(i-k)). This behavior is consistent with the Fermi surface expected for a three-dimensional Dirac semimetal that has approximately equal Fermi velocity in both directions perpendicular to the cleavage plane. The observed behavior is consistent with the Fermi velocity extracted from LL measurements, but the limited number of energies explored precludes more quantitative comparisons.

We also performed extensive STM measurements on a Na$_3$Bi(2) crystals grown from the Na flux. The characteristic hexagonal platelet shape of the high-quality crystals made these samples easy to cleave along the (001) plane, providing a second surface to measure. Figure 3(a) shows a differentiated topographic image with multiple terraces separated by step edges about 4.9 Å in height (Fig. 3(b)). This is in very good agreement with half the *c*-axis lattice constant.

The derivative of the topography highlights the significant surface roughness that we observe in this sample, which may be associated with the high mobility of surface Na atoms or their loss from the surface, as was reported in ARPES measurements [10].

Though the cleaved surface may have disorder associated with Na vacancies or other defects, we observe well-resolved peaks in the density of states when we apply a large magnetic field perpendicular to the surface (Fig. 3(c)). In these higher-quality $Na_3Bi(2)$ crystals, the visibility of the oscillations is more pronounced and the energies of the peaks show only slight position dependence, although the background conductance and the detailed peak shape do exhibit some variability, likely related to disorder on the surface. The differential conductance measured at one location is shown as a function of magnetic field in Fig. 3(d). The peaks in the density of states disperse with magnetic field, and we again associate an orbital index $N$ to each peak. As for the other surface, the peak energies scale linearly with $k_N$ (Fig. 3(e)). From this surface of a flux-grown crystal, we extract a similar Fermi velocity, $v_F$ = 5.05 eVÅ and we find that the Dirac point occurs significantly closer to the Fermi energy: $E_{DP}$ = -16 meV. The increased LL visibility, larger spatial homogeneity and smaller amount of doping all show that the flux grown $Na_3Bi(2)$ crystals are of higher quality than the stoichiometric-melt grown $Na_3Bi(1)$ crystals.

As shown in the resistance vs. temperature curves (Fig. 4(a & b)), both the crystals exhibit metallic behavior. However, $Na_3Bi(2)$ has a smaller resistivity than $Na_3Bi(1)$, and the two types of crystals have different magnetoresistance (MR) profiles. Neither of the two samples exhibits a Drude-like MR profile, possibly due to an unconventional field tuned scattering process in these topological Dirac semimetals (Fig. 4(c & d)). The MR of $Na_3Bi(2)$ is almost linear in field, and the ratio $(\rho(9T) - \rho(0))/\rho(0)$ is approximately 460%, which is much larger than the 50% of $Na_3Bi(1)$. This larger MR ratio in $Na_3Bi(2)$ may be related to a higher mobility in the flux-grown crystals. The higher mobility was confirmed by the observation of clear Shubnikov-de Haas (SdH) oscillations in $Na_3Bi(2)$, while the $Na_3Bi(1)$ crystals studied did not exhibit any quantum oscillations in the MR. Figure 4(e) shows the periodic behavior in the MR signal when plotted as a function of 1/B after a smooth background subtraction. We can easily distinguish more than 30 oscillations starting from 4T, consistent with high crystal quality. Figure 4(f) shows the Landau index plot of the SdH oscillations, which is linear. The SdH oscillation period is given by $2\pi(n + \gamma) = S_F \frac{\hbar}{eB}$ [35], where $S_F = \pi k_F^2$ is the Fermi surface area, $\gamma$ is the Onsager phase and $k_F$ is the Fermi wavevector. From the Landau index fit, we obtain the Fermi surface area, $S_F = 217$ T, corresponding to a Fermi wavevector $k_F = 0.083$ Å$^{-1}$. Hall measurements show that the Na$_3$Bi(1) crystals are *p*-type whereas Na$_3$Bi(2) crystals are *n*-type (data not shown). The details of those transport measurements will be reported elsewhere [36]. In low temperature resistivity measurements, some of the stoichiometric-melt grown Na$_3$Bi(1) crystals exhibited superconductivity ~2 K, corresponding to the superconducting T$_c$ of NaBi [26]. The drop in resistivity as well as the additional peaks in X-ray diffraction reveal the co-crystallization of NaBi as an impurity in stoichiometric-melt grown Na$_3$Bi(1).

Although the STM data come to the same qualitative conclusion as prior ARPES studies that Na$_3$Bi is a 3D Dirac semimetal, there are quantitative differences in the extracted Fermi velocity. The STM data clearly show a linear dispersion with $v_F \approx 5$ eVÅ, whereas ARPES measurements indicate a Fermi velocity of 1-3 eVÅ in the directions perpendicular to the *c*-axis. The STM data agree with the Fermi velocity extracted from transport measurements on the same crystals. One possible explanation for this discrepancy is a difference in the Na$_3$Bi crystals used. In particular, variability in the samples grown by cooling a stoichiometric melt is likely, we have found that Na$_3$Bi(1) crystals taken from the same batch shows regions that are n-type, or p-type, or include the impurity phase NaBi. The kinetics of crystal growth through the high temperature part of the phase diagram appears to give rise to the inhomogeneity of these samples. The Na-flux growth of Na$_3$Bi, on the other hand, gives significantly more consistent results. Alternatively, the apparent difference between the APRES data and the STM data may be due to differences in the measurement techniques. Because the STM measurements show LLs only in the conduction band, and ARPES is only sensitive to occupied states in the valence band, the deduced Fermi velocities can be different if the bands in Na$_3$Bi are electron-hole asymmetric. Finally, we note that we extract a similar $v_F$ on both surfaces. This is somewhat surprising because the Fermi velocity along the [001] direction is predicted to be significantly smaller than the other two directions.

Fig. 5(a) shows a typical core level spectrum of our Na$_3$Bi samples. Both the Na 2*p* and Bi 5*d* core levels are observed, which provides a spectroscopic confirmation of the chemical composition. We study the low energy electronic structure at the (100) surface, where topological surface states due to the nontrivial topology in Na$_3$Bi are predicted [9]. Our previous ARPES measurements showing the 3D Dirac bulk band structure [12], were taken on the Na$_3$Bi(1) batch of samples. Fig. 5(b) shows a schematic drawing of the Na$_3$Bi (100) Fermi surface according to the theoretical calculation [9], which consists of two bulk 3D Dirac nodes and two surface state Fermi arcs connecting the two bulk nodes. The ARPES-measured energy dispersion cuts along the dotted line in Fig. 5(b) are shown in Fig. 5(c) for both the Na$_3$Bi(1) sample and the Na$_3$Bi(2) sample. For Na$_3$Bi(1), only bulk valence bands are found in the ARPES spectrum and the Fermi level is seen to cut within the bulk valence band. By contrast, the ARPES dispersion data for Na$_3$Bi(2) reveals Dirac (linearly dispersive) surface states that cross the Fermi level. The observed surface states are consistent with the theoretical calculation results [9]. Moreover, systematic ARPES data for these surface states are reported in Ref. [30]. The bulk valence bands, on the other hand, are found to be pushed down to higher binding energies. From our ARPES data in Fig. 5(c), the chemical potential of Na$_3$Bi(1) is observed to be lower than that of Na$_3$Bi(2) by about 150 meV. Although ARPES is in general surface sensitive whereas transport measures the bulk properties, the qualitative observation that the Na$_3$Bi(2) sample is relatively more *n*-type than the Na$_3$Bi(1) sample is consistently found in both our ARPES and transport data.

**Conclusions**

Bulk single crystals of Na$_3$Bi have been grown using two different methods: stoichiometric-melt growth and Na-rich flux growth. The flux grown crystals are found to be of higher quality, and possess a clear hexagonal morphology. An NaBi impurity phase has been observed in stoichiometric-melt grown crystals, which also have a less well defined crystal morphology and greater variability in carrier type. The STM measurements confirm the 3D TDS nature of Na$_3$Bi. The LDOS in a magnetic field provided values for the Dirac point energy ($E_{DP}$) and Fermi velocity ($v_F$), of the samples. For Na$_3$Bi(1), $E_{DP}$ = -71 meV and $v_F$ = 5.22 eVÅ and for Na$_3$Bi(2), $E_{DP}$ = -16 meV and $v_F$ = 5.05 eVÅ. Transport measurements show that the MR of Na$_3$Bi(2) is linear and significantly larger that of Na$_3$Bi(1). The flux grown Na$_3$Bi(2) crystals possess high carrier mobility as evidenced by bulk Shubnikov-de Haas(SdH) oscillations. The Fermi surface area is found to be $S_F = 217\ T$, corresponding to a Fermi wavevector of $k_F = 0.083\ \text{Å}^{-1}$. In contrast, the Na$_3$Bi(1) crystals were not found to display SdH oscillations. The ARPES studies on (100) surfaces revealed the presence of 2D surface states in the Na-flux grown crystals. The results presented here demonstrate the high quality of Na$_3$Bi crystals grown by the Na flux method, making it possible to perform measurements which provide strong evidence for the 3D TDS nature of the material.

**Acknowledgements**

This research was supported by the ARO MURI on topological insulators, grant W911NF-12-1-0461, ARO grant W911NF-11- 1-0379, the MRSEC program at the Princeton Center for Complex Materials, grant NSF-DMR-0819860 and grant DOE DE-FG-02-05ER46200.

**Figure Captions;**

**Fig. 1: Crystal growth and structure analysis.** (a) the photograph of stoichiometric cooled "Na$_3$Bi(1)" crystals with naturally cleaved shining faces, (b) Photograph of typical 90% Na flux grown "Na$_3$Bi(2)" crystals (c) the schematic represents the honeycomb crystal structure of Na$_3$Bi, Na(2) atoms are cross linked to Bi of -Na(1)-Bi-Na(1)-Bi hexagonal layers (d) the schematic representation of hexagonal morphology of flux grown crystals, arrow indicates the c-direction [001]. (e) the recorded X-ray diffractograms, upper (blue) for powder specimen of Na$_3$Bi(1) crystals and lower (orange) for Na$_3$Bi(2) single crystal. The powder pattern for Na$_3$Bi(1) is in agreement with reported ICSD data (card number 26881), plotted in green vertical lines. The additional peaks in the powder pattern indicated by red stars correspond to an NaBi impurity phase. The observed peaks for the Na$_3$Bi(2) crystal correspond to the (00l) reflections, which confirms that grown crystal is single crystalline block. The inset in (e) shows a schematic

for the Na-Bi phase diagram Ref. [29]; the orange shaded region indicates the crystallization of Na$_3$Bi with wide a composition range and how NaBi may appear as an impurity. The vertical dashed orange line indicates the Na-rich composition used for flux crystal growth.

**Fig. 2: STM measurements of Na$_3$Bi(1).** Topographic images of (a) 100 nm x 100 nm and (b) 15 nm x 15 nm areas on the cleaved surface of stoichiometric-melt-grown Na$_3$Bi (1) at $V_{bias}$ = -500 mV and $I$ = 30 pA. The Fourier transform of the topography (inset) shows the modulation associated with the one-dimensional stripes (orange circles), and the quasi-periodic structure (red circles). (c) Line cut of a region with single and double atomic steps displaying an average step height $\Delta z \approx 4.6$ Å. (d) Point spectra taken at a single location on the surface similar to the one shown on Fig. 1a obtained at high magnetic fields from 10 to 14 T. The peaks in the (local density of states) LDOS correspond to the formation of Landau levels (LL; $V_{bias}$ = - 100 mV and $I$ = 100 pA). The individual spectra are offset for clarity. Indices correspond to the LL assignment. (e) Energy of each LL as a function of momentum reveals a linear relationship. The black solid line is a linear fit to the data. The real space $dI/dV$ conductance maps (f-h) at different energies on the surface shown in (a), and their corresponding Fourier transforms (i-k). Data were acquired at $V_{bias}$ = 600 mV parking bias and $I$ = 500 pA setpoint current. The STM measurements were performed at a temperature of approximately 3K.

**Fig. 3: STM measurements of Na$_3$Bi(2).** (a) Differentiated topographic image of the (001) surface of Na-flux-grown Na$_3$Bi(2) ($V_{bias}$ = -500 mV and $I$ = 30 pA). The red line indicates the position of the line cut in (b), which reveals step heights of $\Delta z$ = 4.9 Å corresponding to half of the c-axis lattice constant. (c) Spatial variability of the LDOS was observed at the 13 T high magnetic field. The location of each spectrum is indicated in (a). (d) Landau levels as a function of magnetic field measured on the (001) surface at a single location ($V_{bias}$ = -200 mV and $I$ = 100 pA) (e) Energy-momentum structure of the LL peak positions shows a linear dispersion relation. The black solid line is a linear fit to the data. The STM measurements were performed at a temperature of approximately 3K.

**Fig. 4: Transport measurements.** (a) & (b) Resistance vs. temperature curves of Na$_3$Bi(1) and Na$_3$Bi(2), showing metallic behavior. (c) & (d) are the magnetoresistance, Na$_3$Bi(1) has a smaller MR ratio, possibly due to lower mobility. (e) The Shubnikov-de Haas oscillations in MR of Na$_3$Bi(2), a smooth background has been subtracted to emphasize the oscillation signal. (f) The Landau index fan diagram of SdH oscillations in Na$_3$Bi(2). The red circles are the maxima of the resistance, while the blue triangles are the minima of the resistance and the green line is the linear fit.

**Fig. 5: ARPES measurements on Na$_3$Bi.** (a) Core level spectroscopy measurement on a Na$_3$Bi(1) sample using incident photon energy of 90 eV. Both the Na 2$p$ and Bi 5$d$ core levels are observed. (b) A schematic drawing of the Na$_3$Bi Fermi surface on the (100) surface according to the theoretical calculation in Ref. [11]. The Fermi surface consists of two bulk Dirac nodes and a pair of Fermi arc surface states (SS) that connect the bulk nodes. The black dotted line denotes the momentum space cut-direction for the energy dispersion maps shown in panel (c). (c) ARPES dispersion maps along the dotted line in panel (b) for the Na$_3$Bi(1) (left) sample and the Na$_3$Bi(2) sample (right). Na$_3$Bi(2) is found to be relatively more $n$-type than Na$_3$Bi(1). BVB = Bulk Valence Band.

Figure 1

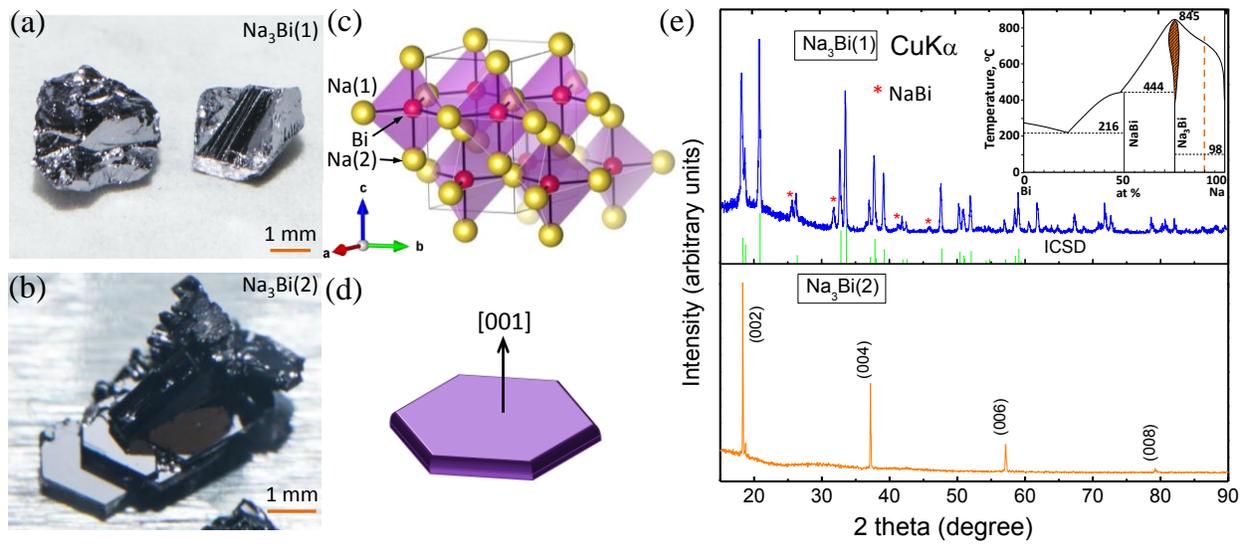

Figure 2

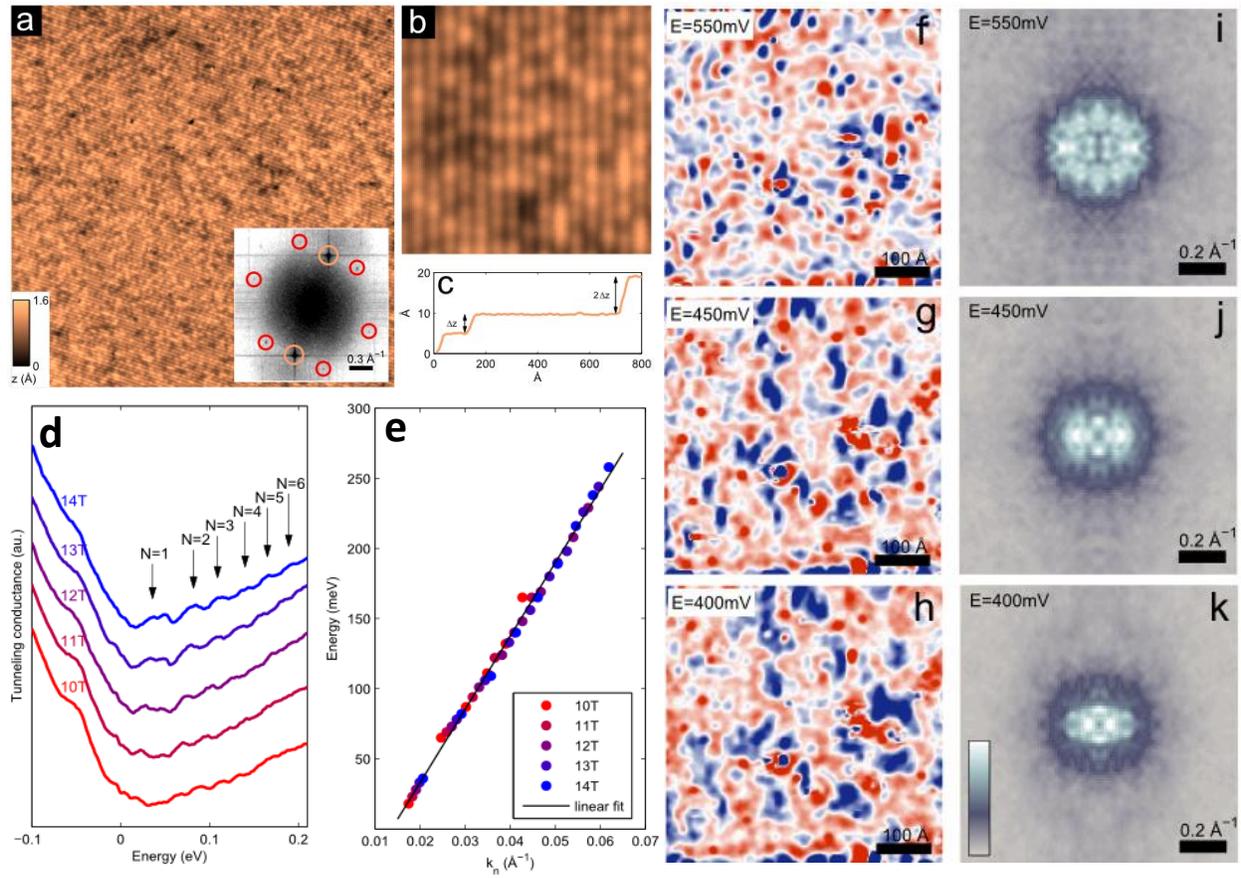

Figure 3

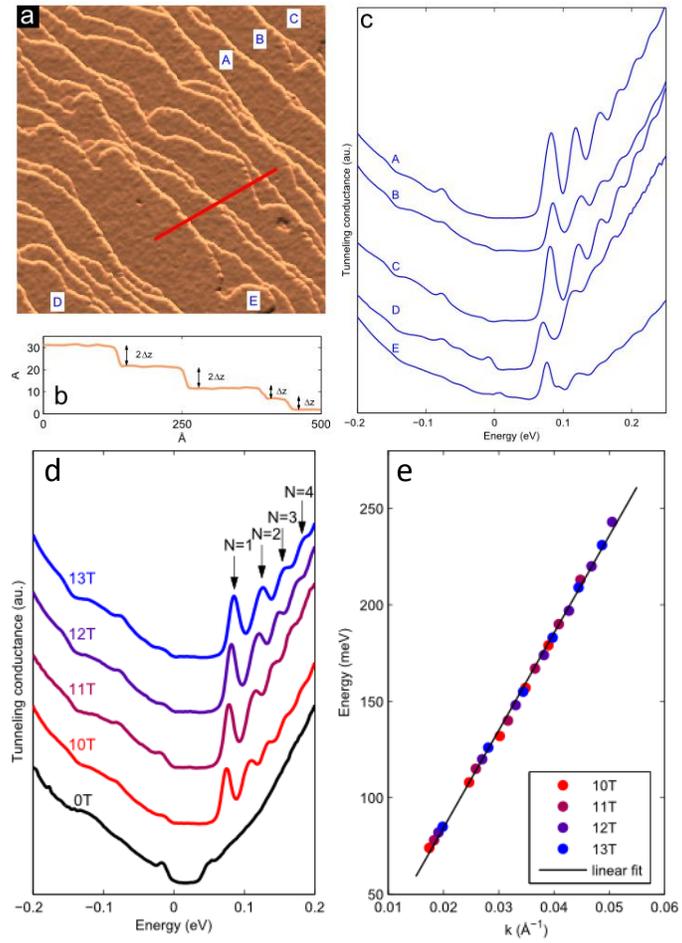

Figure 4

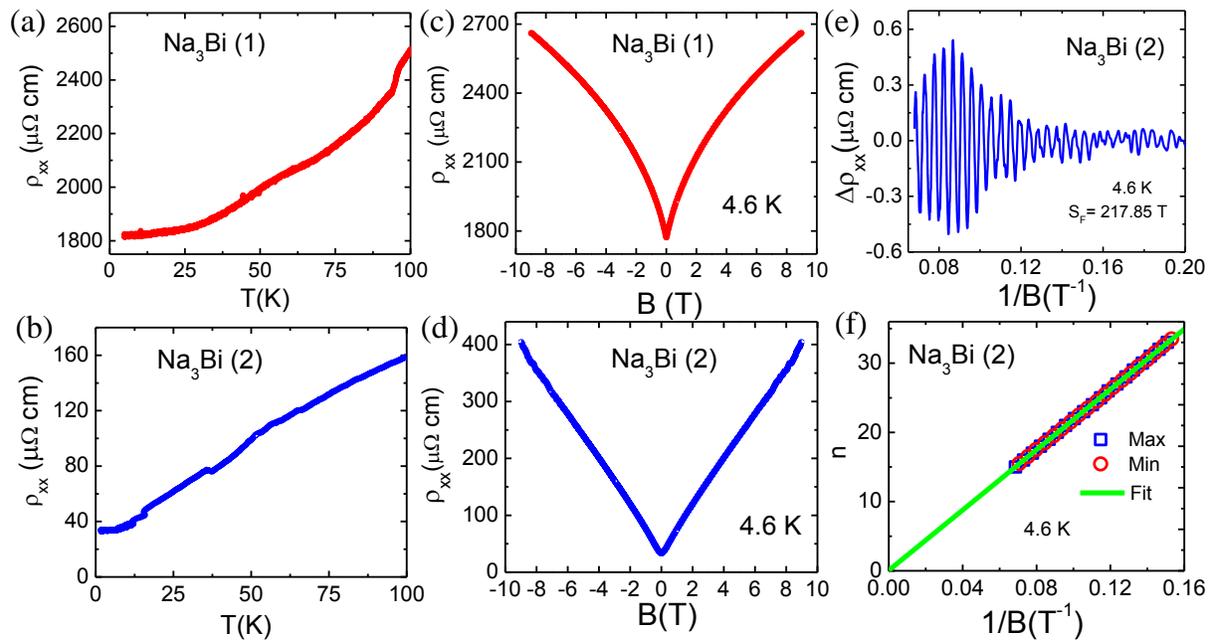

Figure 5

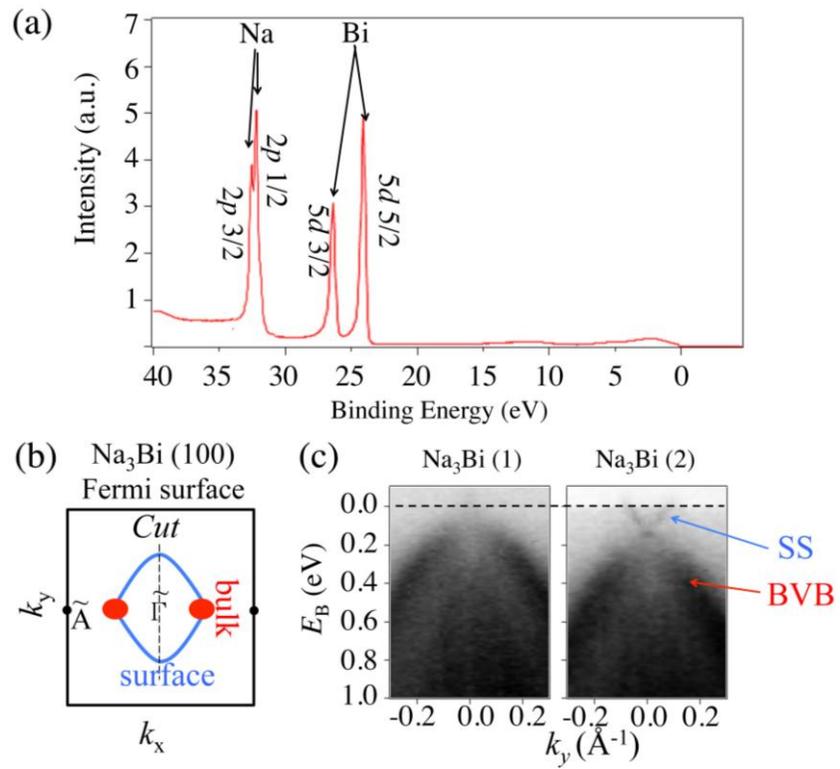